\documentclass[aps,showpacs,preprintnumbers,amsmath,amssymb, 
twocolumn, tightenlines]{revtex4}

\usepackage[dvips]{graphicx}\usepackage{bm}
\usepackage{amsmath}
\usepackage{dcolumn}
\usepackage{bm}

\newcommand{\be}{\begin{eqnarray}}
\newcommand{\ee}{\end{eqnarray}}

\begin{document}
\title{On the Origin of the ``Ridge" phenomenon\\
 induced by Jets in Heavy Ion Collisions}

\author{E.Shuryak}
\affiliation{ 
Department of Physics and Astronomy, State University of New York, 
Stony Brook NY 11794-3800, USA
}

\date{\today}
\begin{abstract}
We argue that ``ridge" in 2-particle correlation function associated with hard trigger
at RHIC heavy ion collisions is naturally explained by an interrelation
of jet quenching and  hydrodynamical
transverse flow. The  excess particles forming
the ridge are produced by QCD bremsstrahlung along
the beam (and thus have wide rapidity distribution) and then boosted by
transverse flow. Nontrivial correlation between directions of the jet and the radial flow 
is provided by jet quenching: our straightforward and basically parameter-independent
calculation reproduces the angular shape, width and other properties of the ``ridge".
\end{abstract}

\vspace{0.1in}

\maketitle

\section{Introduction}
 Two most important discoveries made in first years
of heavy ion collision experiments at RHIC
are (i)  robust radial and elliptic flows which are
well described hydrodynamically \cite{hydro}, and   (ii) strong
jet quenching. In the last years one of the hot subjects became
an interaction between jets and the medium. Strong modification
of the away-side jets seem to be well described by
another hydrodynamical  ``conical flow''\cite{conical}.

This paper is about another phenomenon observed in jet-related
2-particles correlations, known as ``ridge'' and found by STAR
collaboration. It was originally observed in fluctuation analysis
\cite{Adams:2005aw} revealing ``mini-jets'', and then related to
few-GeV jets,
 for recent summary see e.g. \cite{Putschke}).
Its main features are: (i) a peak at 
relative azimuthal angle $\phi=\phi_1-\phi_2=0$
 with a width
 of about 
1 radian, about twice that of the jet; (ii)  wide
distribution in (pseudo)rapidity $\eta$; (iii) a spectrum of
secondaries slightly harder than a bulk one but much softer than
that for a jet; (iv) a composition very different from jets,
in particular large fraction of baryons/anti-baryons. 
  
 We would not go into a review of various ideas
proposed to explain the ``ridge''.  We just report our calculations
aimed at testing
one specific idea, originating from the paper of S.Voloshin 
  \cite{Voloshin:2003ud}, who  pointed out that
one can get information about the
location of the hard collision point by correlating its with
the transverse flow.  To our knowledge, the present paper is the first
attempt to make quantitative estimates based on it,
with results we consider very encouradging.

\section{Angular correlation between jet and flow}

 As it is well known, radiative QCD processes lead to production
of four cones of radiations. Two of them -- the ``jets'' --
are better  known and studied, since two others
produced along the beams. While they are similar in multiplicity 
and other features to two ``jets'' (because appearance and disappearance
of the same color current produces similar radiation),
the hadrons originating from them  cannot be
separated from ``bulk'' multiple production in pp collisions.
Indeed, they have similarly wide
rapidity distribution, and similar transverse momenta
$p_t$ in respect to  the beam direction, so
their presence may only be seen via overall multiplicity increase in
jet-containing events, relative to ``soft'' ones.

\begin{figure}[t]
   \includegraphics[width=4.5cm]{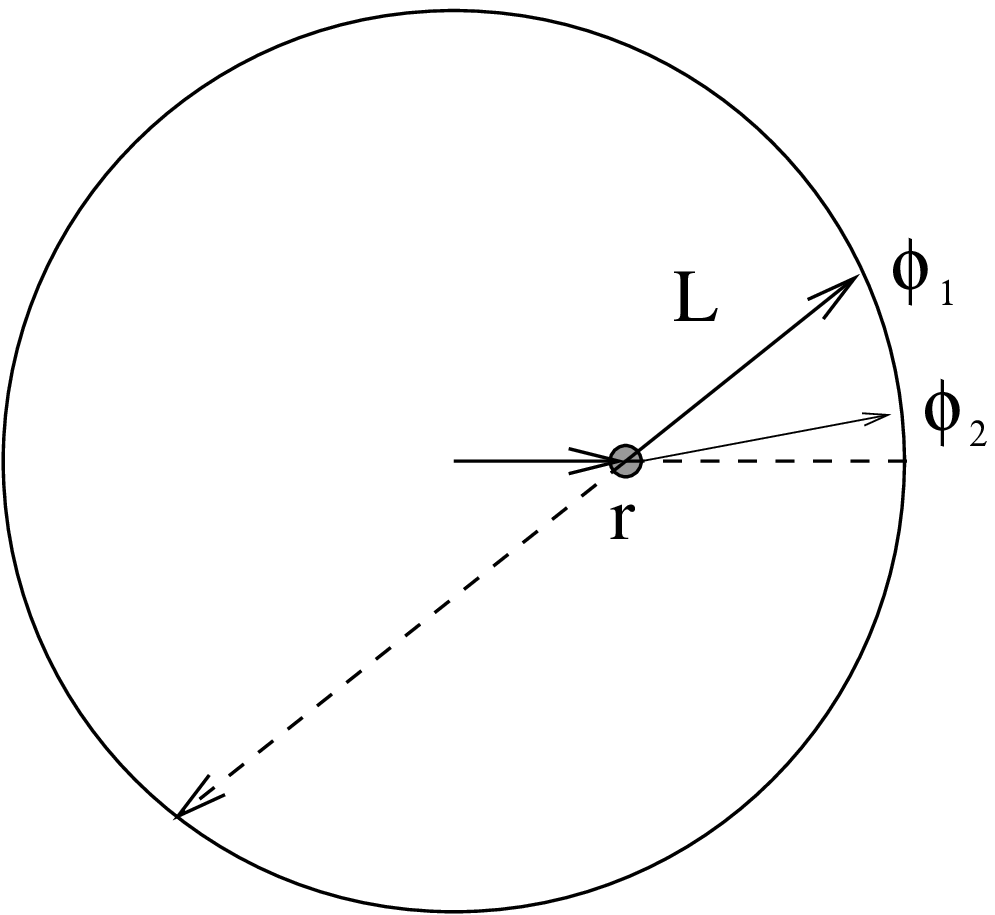}\\
   \includegraphics[width=6.cm]{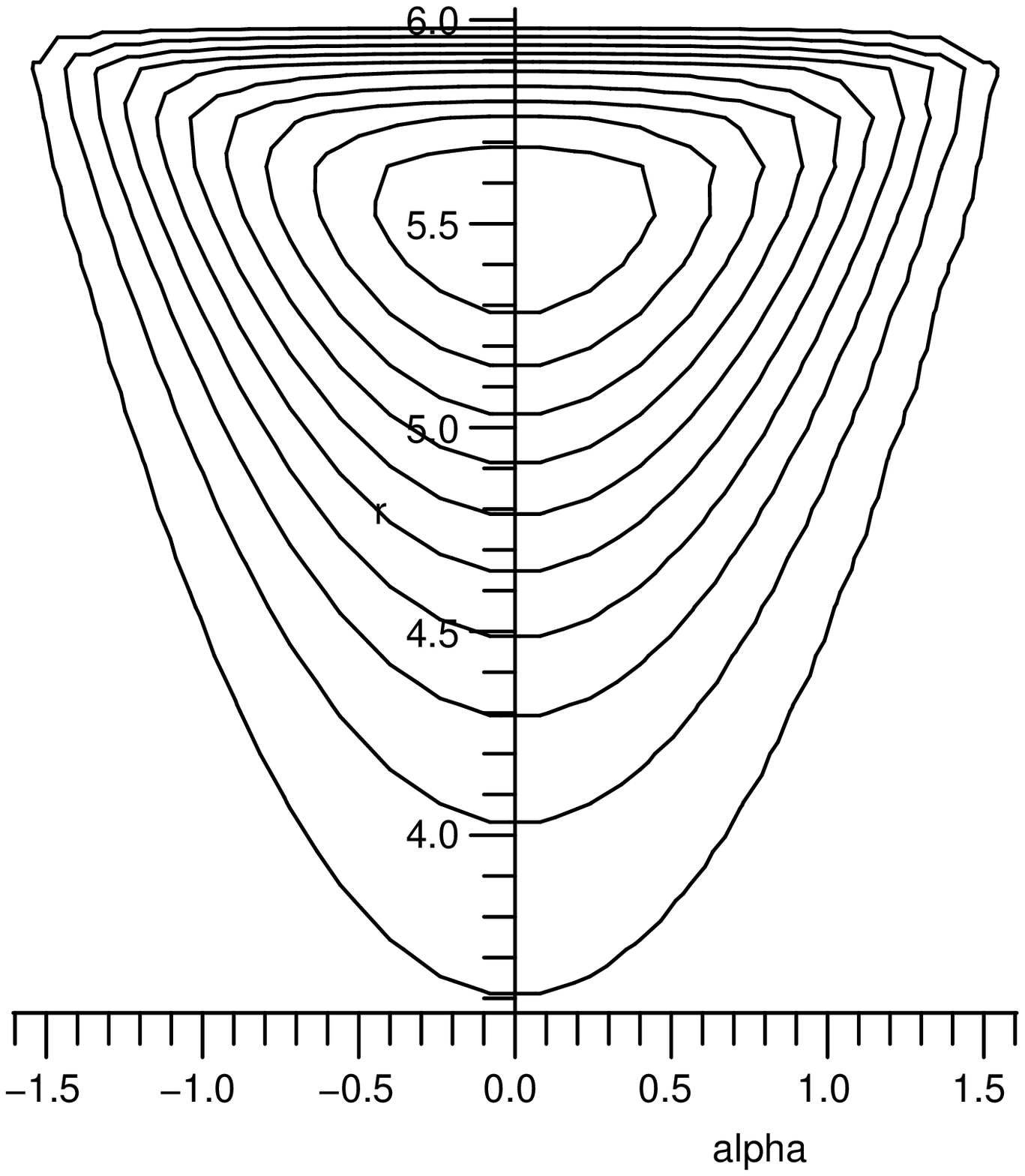}
  \vspace{0.1in}
\caption{(a) 
Schematic view of the transverse plane for central heavy ion
  collisions. The small circle at coordinate $\vec r$ is the place
where hard collision takes place, in  which a pair of jets (triggered
  one shown by a solid line, unobserved by the dashed one) are
  produced. The distance traveled by the triggered jet inside the
nucleus is  $L$, which depends on  $\vec r$ and jet direction angle
$\phi_1$. 
(b) Distribution in $r-\phi_1$ plane.
\label{fig_plane}}
\end{figure}

In heavy ion collisions the situation is different: as we show below,
the ``longitudinal cone'' products
can be naturally separated from the ``bulk''. The  reason for that
is their specific production locations in the transverse
plane -- the gray circle in Fig.\ref{fig_plane}(a) -- which tend
to be closer to the nuclear edge than to the  center,
due to jet quenching. 
Collective  transverse flow boost them strongly in 
 the radial direction, making their azimuthal directions
to be well aligned along  $\vec r$  (especially if one
select the right window of $p_t\sim 2\, GeV$, see below). 
The next step, explaining why this effect is observable,
is a correlation between the radial direction and that of the
triggered
jet. 

The geometry of the phenomenon and the notations used is explained
in Fig.\ref{fig_plane}(a) , depicting transverse plane at the moment of a collision. For simplicity
we discuss only central collisions, for which there is perfect axial symmetry and
the elliptic flow is absent.
  The point at which hard collision takes place is denoted by $\vec r$
and the (azimuthal) angle at which triggered jet is emitted is called $\phi_1$.
At the moment of production obviously there is no correlation between 
directions of $\vec r$ and $\phi_1$. However this correlation appears for $observed$ jets due to jet quenching phenomenon. Indeed, in order to be detected jet has to go through matter
a distance (depicted $L$ at the figure) at which quenching takes place,
the probability of which we call $P_{quench}(L)$. Since obviously the distance
depends on both $r,\phi_1$ and   the nuclear radius $R$  
\be L(r,\phi_1)= \sqrt{R^2-r^2*sin^2(\phi_1)}-r cos(\phi_1) \ee
this generates the correlation between them to be explored. 

Since it is the main
point of the phenomenon, let us discuss  it   in detail.
If a jet is produced at small $r$ 
close to the nuclear center (where the probability of production
$P_{prod}(r)$ has obviously its maximum) there is no correlation, since $L$
in this case is about the same $\approx R$ 
in all directions. If the
jet is produced near the nuclear surface $R-r\ll R$ there is some
angular  correlation, but a weak
one: in this case jet may
  be emitted in the whole half-plane $-\pi/2<\phi_1<\pi/2$.
The correlation reducing $\phi_1$ distribution to more narrow  
peak appears only  when jets
originate at
a certain depth inside the nuclei: and the question to be addressed
is whether it is strong enough
to explain the observed effect. 

Although we have used different variants of distributions in the
 study,
 it is found to be enough to use the simplest models of the
production/quenching , as the results are found
to be  insensitive to any details. For 
central collisions of two homogeneous balls of radius $R$, with a
sharp edge, at position $r$ one has collision of two columns of matter
with a length $\sqrt{R^2-r^2}$, and thus ``collision scaling'' means
\be P_{prod}(r)\sim (R^2-r^2) \ee
The probability of
quenching can be written as a simple exponential damping with distance
\be P_{quench}(L)\sim exp(-{L(r,\phi_1)\over l_{abs}})\ee 
where $l_{abs}$ is the quenching length. 
The resulting distribution $ P_{prod}P_{quench}$ in $r-\phi_1$ plane
is shown in Fig.\ref{fig_plane}(b): one can see that for $r=4-5 fm$
the width of $\phi_1$ distribution is about one radian. This width
would eventually become the observed width of the ``ridge'',
as will show below.

 The next step in the calculation is to address the effect of the
 radial flow on spectra of secondaries. As usual, those are determined
from Boltzmann thermal distribution  at
kinetic freezeout temperature $T_f$, boosted by the flow velocity to 
\be {dN\over dy dp_t^2}\sim exp(-{u_\mu p_\mu\over T_f})\ee
Here the nonzero components
of the flow velocity are written as $u_0=1/\sqrt{1-v^2}, u_r=
 v/\sqrt{1-v^2}$ (because we focus on the transverse flow).
Since we need anisotropy, we can focus on the second term
in the exponent,
containing the angle $\phi_2$ between the particle 2 and
the flow direction:
\be 
F(p_t,v,\phi_2)= exp({ v p_t cos(\phi_2) \over \sqrt{1-v^2} T_f})\ee
To get a feeling of the degree of collimation, let us
estimate of the combination of parameters entering
this exponent. We take $p_t\approx 2.25 \, GeV$ (the lowest
$p_t$ used by STAR
in ridge studies to be discussed below) and 
 $T_f\approx 100\, MeV$. At the edge
of the fireball  $v\approx 0.7$ and thus a distribution
$F\approx exp(-11\phi_2^2))$ which is extremely well collimated,
 with a width much less than that of the
observed ``ridge''. At the  opposite limit, at the center $r=0$,
there is no radial flow, $v=0$, and $\phi_2$ distribution is isotropic.

\begin{figure}[t]
   \includegraphics[width=6.cm]{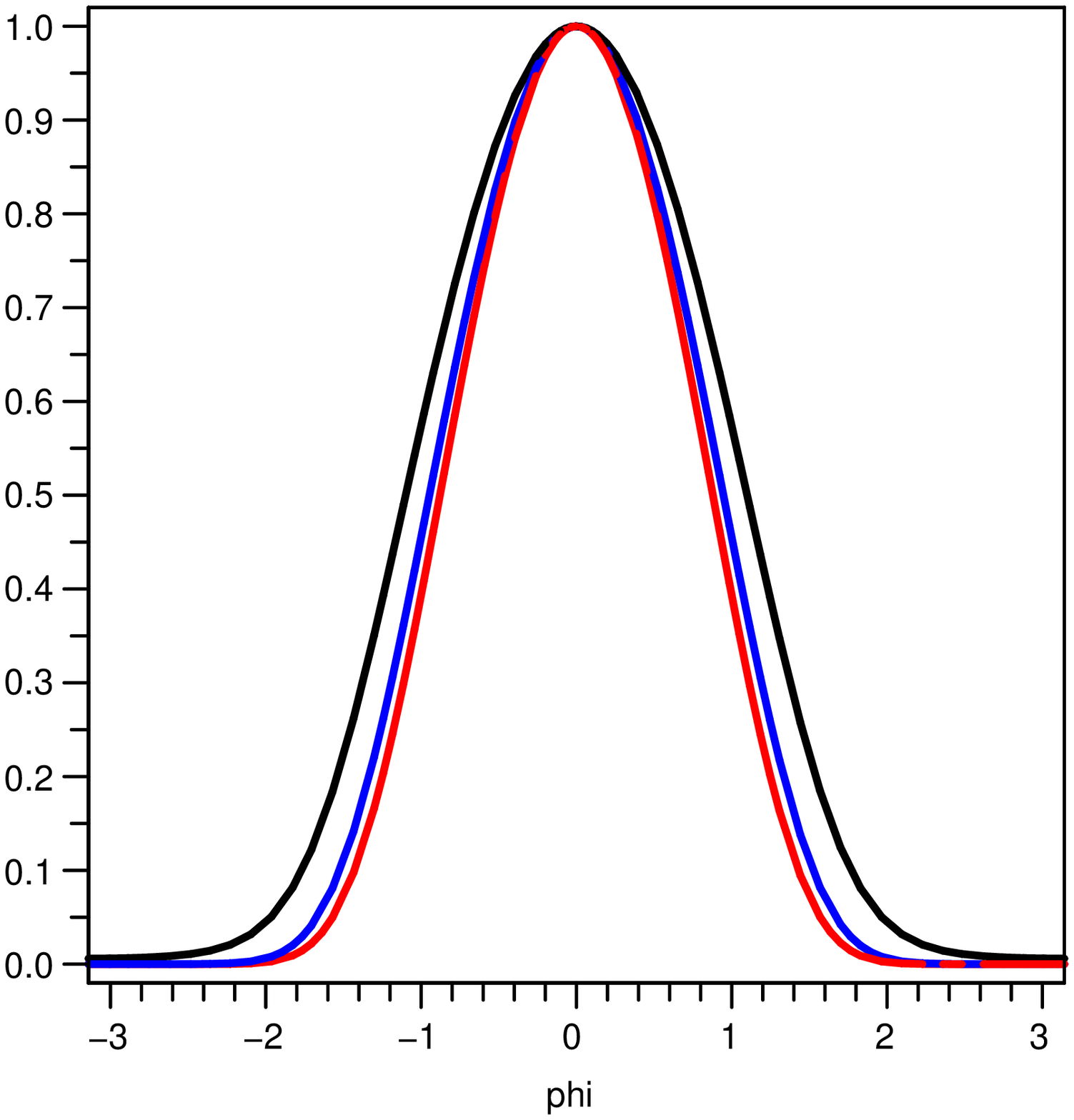}\\
   \includegraphics[width=7.cm]{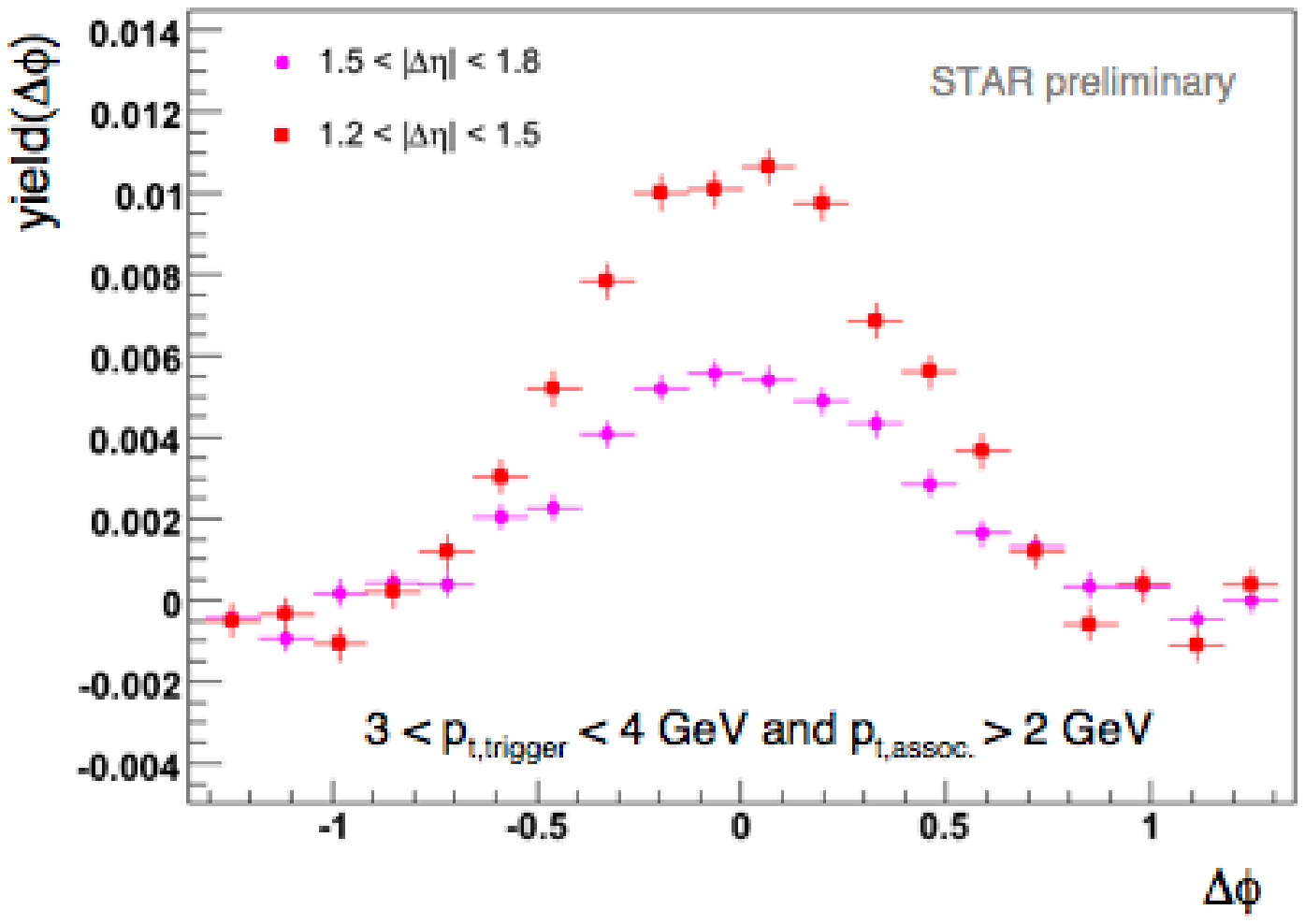}
  \vspace{0.1in}
\caption{
(a) Calculated correlation function $C(p_t=2.25\, GeV,\phi)$ 
as a function of the angle between two observed particles.
Three curves (top to bottom) correspond to absorption length
$l_{abs}=1,0.5,0.25 \, fm$.
(b)Preliminary STAR data \cite{Putschke}
 on the ``ridge'' shape as a function
of  $\phi$, with the jet component subtracted. AuAu at $\sqrt{s}=200\,
GeV$ and the central collision bin 0-10\%.
\label{fig_phi}}
\end{figure}

Thus the remaining task  to be performed is the averaging over both
the jet origination point $r$ and  the angle $\phi_1$, with
the weights given by the distributions discussed
 above. Furthermore, the experimentally  observable  angle 
is neither $\phi_2$ nor $\phi_1$ but the angle 
$ \phi=\phi_1-\phi_2 $ 
between 
particles 1 and 2, and so
the correlation function is
\be C(p_t,\phi)= \\
\int P_{prod}(r) P_{quench}(r,\phi_1)
F(p_t,v(r),\phi_1-\phi) r dr d\phi_1 \nonumber
 \ee
The only remaining input needed is the ``Hubble law''
for the radial flow, which we use in the form\footnote{This expression
is supposed to give the final velocity which is obtained by
the volume element which started
hydro expansion at the point $r$. It should not be confused with
a solution of hydro equations at some intermediate time moment.
}
\be v(r)=r/(10\, fm)\ee
In Fig.\ref{fig_phi}(a) we show the resulting 
angular distributions: the main result is that the peak 
survives the averaging. Furthermore, for small enough 
absorption lengths $l_{abs}$ shown the result is remarkably
independent on it: and since the absorption length
is the only parameter of the model, and is believed to be rather
small, we call the calculation ``parameter free''. Still the reader
should be warned that for weak quenching
 $l_{abs}>3 fm$ the width of the $\phi$ distribution grows
catastrophically and the ``ridge'' correlation disappears.

While
comparing these distribution to STAR data (Fig.\ref{fig_phi}(b)) one finds
that the model is $not$ quantitatively accurate: the
width we found is  larger than the one observed.
 By making more complicated models for quenching one probably
can recover better agreement: all we conclude for now is that
the mechanism of ``ridge'' formation basically works.

\section{Other observables}
{\bf Spectra} of particles belonging to a ridge
 are very different from those
of the jet, being  much softer. 

\begin{figure}[t]
   \includegraphics[width=10.cm]{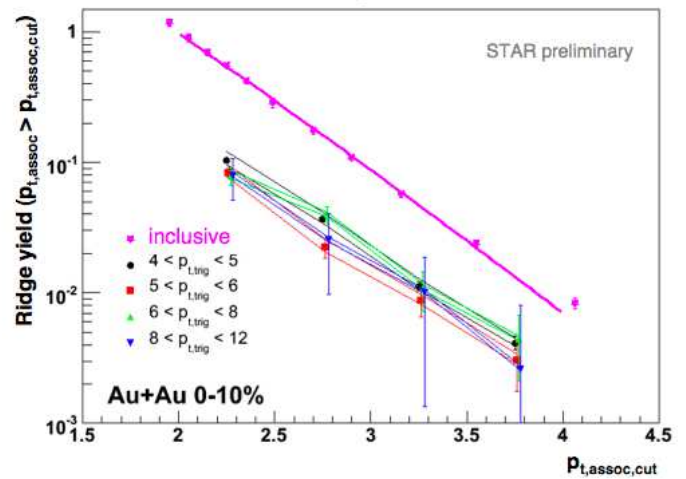}
  \vspace{0.1in}
\caption{Preliminary STAR data \cite{Putschke} on the spectra
of the particles originating from a ``ridge'', for different triggers.
Triangles show spectrum of the ``bulk'' inclusive spectrum
in the same momentum window.
\label{fig_spect}}
\end{figure}

As
one can see from
Fig. \ref{fig_spect}, they are much closer to the ``bulk" represented by inclusive
spectra. In fact they have a bit stiffer slope, which is naturally explained
by the fact that  the distribution over their points of origin $r$
discussed above is more biased toward the nuclear surface, than for the
bulk  matter. 

Another important conclusion from Fig. \ref{fig_spect} is that the
spectra are completely independent on the jet momentum, which
confirms
that the ``ridge'' is not physically related to a jet itself.
This fact is consistent with our model, since different jets have
the same ``collision scaling'' distribution in the transverses plane.

{\bf Particle composition} of the ridge particle is also very different from that of the jet.
The fraction of baryons is much larger. This is naturally explained by the fact
that the ridge is seen in the region of $p_t\sim 2 \, GeV$ which constitute
the tail of (boosted) Boltzmann distribution in which mass dependence is small.
The same very phenomenon was observed in the bulk, ad was explained by
hydrodynamics \cite{hydro}. Indeed, around $p_t\sim 2\, GeV$
the $p/\pi^+$ ratio crosses 1, and if the hydro-induced
 tail would dominate the spectrum
at arbitrary large $p_t$ (which it is not)
 the ratio would eventually
be mass independent and reach
2, the number of  spin components.

\section{Outlook: 3-particle correlations}
The next step in data analysis is obviously adding one more 
particle correlated with the jet. Depending whether
the second particle is included in the trigger condition
or not, those can be called (2+1) or (1+2) correlations.

The latter case is basically the same as (1+1) in terms of
geometry and trigger bias. In this case
  one would like to 
check whether the ridge extends longitudinally
 on both sides from a jet in each event. The alternative mechanism
suggested in \cite{Armesto:2004pt}
 -- a longitudinal extension of a jet due to
$longitudinal$ flow -- can thus be finally confirmed or rejected.
So far, the only observation $against$ it is that the ridge was never
seen near the away-side jet, which their model seem to predict 
to be even larger than the observed ridge at the trigger side.

The (2+1) case, with two hard particles in the trigger, is
completely symmetric if two momenta are about the same, and
therefore its trigger bias is  completely different from the one
discussed above.
Indeed, it is determined by quenching
along the sum of the paths
of both jets
\be L+\bar L= 2\sqrt{R^2-r^2*sin^2(\phi_1)} \ee
where $\bar L$ is the path of companion jet shown in Fig.1 by the dashedline.
Its exponent now favors the flow vector $\vec r$ to be
 $orthogonal$ to both jets,
$\phi_1=\pm \pi/2$.
The favorite configurations is when two jets are emitted 
``tangentially'' to flow: therefore we predict
 that now one should find the ridge at a completely different location!
In rapidity it is expected to be 
symmetric around   the  di-jet center-of mass,
 the mean
of the rapidities of both jets.

\section*{ Summary}

 In short, the proposed mechanism works as follows. The ``ridge''
particles originate from glue radiated in the hard collision
along the beam direction, with calculated angular collimation coming
from transverse radial flow. The most nontrivial point is the
correlation between the
direction of the flow and jet direction, which
is induced by the jet quenching: as we show, it
survives the averaging over positions and jet directions.
We conclude that  this mechanism 
  is in good correspondence with many aspects of the data
on the ``ridge'' phenomenon at hand.
Further experimentation, especially with 3-particle correlations,
will further elucidate whether this mechanism is indeed
responsible for this phenomenon.

\section*{Acknowledgment}

My interest to this issue stems from discussions
during  the workshop 
organized by B.Cole and J.Dunlop (2007 AGS/RHIC users meeting,June
19 2007). I also acknowledge  email comments from J.Putschke.
This work is  supported by the US-DOE grants DE-FG02-88ER40388
and DE-FG03-97ER4014.

\end{document}